# Data Breach e-Crime, A case study, and Legal Analysis Bonafede V. EE

Samuel Cris Ayo
u1815973@uel.ac.uk

## Abstract

The *Bonafede V. EE data breach* is a reported data breach e-Crime in the media, also published by the BBC on 8[th] February 2019 in the United Kingdom which has not yet come to Court. Three laws and regulations of the United Kingdom that have been breached and references to relevant case law are highlighted in this paper. Further, Facts of the Case; Issues observed; Decision made; Reasonings to the case; Opinions; and Analysis are discussed. The discussions include legal points raised in the case, with the relevant laws to draw attention to the keywords or phrases that are in dispute. There are sufficient details of the organization for the reader to understand both the scale and its activity. Discussions refer to case law in support with the reasonings outlined, point by point in numbered paragraphs. At the analysis stage, the significance of the case, its relationship to other referenced cases, its place in history and the value of the current laws in place are evaluated while making references to current law.

## Definitions
*"Computer", "cyberspace", "e-Crime" and "cybercrime"*

(1) The terms "e-Crime" and "cybercrime" are often used interchangeably (Home Affairs Committee, 2013) (Ian, 2016). It is talked about whether relating to information theft, hacking or denial of service to vital systems as a becoming fact of life (Ian, 2016). However, before e-Crime (or cybercrime) can be defined, there is a need to know some terms that are commonly used in relation to e-Crimes, which are "computer" and "cyberspace".

(2) A "computer", according to (Collin, 2004) is defined as any machine that receives or stores or processes data very quickly using a program kept in its memory.
The "cyberspace" on the other end is the world in which computers and people interact, normally via the internet (Collin, 2004). However, the term "cyberspace" as stated by (Susan & Gary, 1999) can also be used to encompass a variety of forms of computer-mediated communication ranging from mass communication to interpersonal interaction.

(3) It should be recognized that different professional organizations and bodies have variations in their definitions of "e-Crime" based on how it is understood within their context of use but the most outstanding definitions shall be discussed as below.

(4) "e-Crime" is defined by the police as the use of any computer network for crime (Home Affairs Committee, 2013) (Diane, et al., 2017). The Association of Chief Police Officers (ACPO) defines "e-Crime" as any form of anti-social behavior over the internet or via a mobile device. It is an attack or abuse, using technology, which is intended to cause another person harm, distress or personal loss (Association of Chief Police Officers, 2018). ACPO also uses the following definition of "e-Crime" in its 2009 e-Crime Strategy as "the use of networked computers or internet technology to commit or facilitate the commission of crime" (Association of Chief Police Officers, 2009) (Home Affairs Committee, 2013).

(5) Other bodies and organizations like the Council of Europe's Cybercrime Treaty use the term "cybercrime" (or "e-Crime" in this case) to refer to offenses ranging from criminal activity against data to content and copyright infringement. The United Nations Manual on the Prevention and Control of Computer-Related Crime includes fraud, forgery, and unauthorized access with its definition of cybercrime (Home Affairs Committee, 2013).

(6) From the above definitions buildup, e-Crime can simply be defined as the use of any computer in the cyberspace to commit a crime.

## Facts of the Case
*Overview*

(7) This is a privacy and personal information data breach e-Crime case involving a data subject Ms Francesca Bonafede ("Bonafede") who is a EE customer, her ex-partner (a third-party), the data controller EE as recently reported in the media including the BBC website with the headline " *EE data breach 'led to stalking'* " (BBC, 2019). It is a very recent data breach e-Crime committed in the month February of the year 2019 which has not yet been presented in courts as is still under investigations in the United Kingdom ("UK") as of the month of March in the year 2019.

*Details of the organization*

(8) EE is a division of the BT Group, a British multinational telecommunications holding company headquartered in London, UK which runs the largest and most advanced mobile network in the UK. According to its official website, EE boasts of connecting over 35 million customers in the UK with a 50 percent presence in all households. In terms of scale and interactions, EE reports having 620 retail stores, 1 billion customer contacts per year, 1.2 billion web visits per year and 200 million app visits per year (EE, 2018).

*The background to the Proceedings and Issues between the parties*

(9) For five days, EE customer Ms Bonafede's phone lost network signal and when she contacted EE in February 2018, she was informed someone came in and requested a new SIM (Subscriber Identification Module) replacement then switched the account to a new handset. From the address details recorded, that someone was later identified as her ex-partner. All her personal details including address and bank details were accessed unlawfully. Ms. Bonafede thinks it is also probable that all her text messages and calls made during the five days her phone lost signal would have gone to her ex-partner.

(10) On how she was handled by EE when this breach was discovered, Ms. Bonafede responded "The Agent just didn't seem concerned at all" referring to the EE employee that attended to her. Escalation attempts to engage senior management in EE was frustrated by the same employee. EE failed to respond seriously until the Police were later involved in the case.

(11) Ms Bonafede's ex-partner had called and texted her "endless times" before to persuade her to withdraw a complaint Ms. Bonafede had reported about him. He even showed up unannounced at her new address with his friends multiple times. On reporting to Police, he was arrested and given harassment warnings. Because of the stalking, "endless texts" and unannounced show up of her ex-partner at her new home address, Ms. Bonafede was greatly distressed.

(12) EE only started taking the case seriously when Ms. Bonafede started publicly tweeting about the problem, even after EE's earlier assurance to investigate the matter. EE also acknowledged its internal policies were not followed and apologized to Ms. Bonafede accordingly. EE further added that following the incidence, the employee no longer worked with them. The breach incidence was later reported to the Information Commissioner's Office (ICO) after it was confirmed and is currently being investigated.

## Issues
*Legal points raised and associated law*

(13) EE failed to protect the personal data and identify of their customer Ms. Bonafede from unauthorized access by a third party. By EE processing and switching subject's data to a new handset as presented in *paragraph (9)*, the data subject was not protected against the unlawful use of her personal data by a third-party, her ex-partner. This is a direct violation of the European Union General Data Protection Regulation (GDPR) Article 5 section 1(f) and section 2 of Principles relating to the processing of personal data (GDPR, 2018):

*"1 (f). Personal data shall be processed in a manner that ensures appropriate security of the personal data, including protection against unauthorized or unlawful processing and against accidental loss, destruction or damage, using appropriate technical or organizational measures ('integrity and confidentiality').*

*2. the controller shall be responsible for, and be able to demonstrate compliance with, paragraph 1 ('accountability')"*

(14) Even with a contract for service (that is the phone) between the parties, EE still went ahead to process and share data subject's SIM , address and bank details ( *paragraph (9)* ) with a third-party who is not the rightful owner of the personal data without the verified consent of the data subject which is a breach of

    i.    Section 2 (1) (a) (b) of the Data Protection Act 2018 Protection of personal data (Data Protection Act, 2018)

2. *"(1) The GDPR, the applied GDPR and this Act protect individuals with regard to the processing of personal data, in particular by—*

*(a) requiring personal data to be processed lawfully and fairly, on the basis of the data subject's consent or another specified basis,*

*(b) conferring rights on the data subject to obtain information about the processing of personal data and to require inaccurate personal data to be rectified"*

    ii.    and breach of European Union General Data Protection Regulation (GDPR) Article 6 section 1 (a) (b) of Lawfulness of processing (GDPR, 2018) which states:

*"Processing shall be lawful only if and to the extent that at least one of the following applies:*
*(a) the data subject has given consent to the processing of his or her personal data for one or more specific purposes;*

*(b) processing is necessary for the performance of a contract to which the data subject is party or in order to take steps at the request of the data subject prior to entering into a contract;"*

(15) In *paragraph (10)*, it is noted that contact attempts by the data subject to senior management (the data controller and data protection officers) when Ms Bonafede was trying to have the data breach addressed was frustrated and denied by the EE employee (the data processor). This is a failure to comply with the data controller's general duties as outlined in the Data Protection Act 2018 sections 44 (1) (a) (b) of Information Controller's general duties (Data Protection Act, 2018):

> *"The controller must make available to data subjects the following information (whether by making the information generally available to the public or in any other way)—(a) the identity and the contact details of the controller; (b) where applicable, the contact details of the data protection officer"*

(16) Because of EE's negligence in *paragraph (9)* which resulted into the data breach exposing personal family details, the Human Rights Act 1998 section 1 article 8 (1) Right to respect for private and family life (Human Rights Act, 1998):

> *"1. Everyone has the right to respect for his private and family life, his home and his correspondence"*

(17) Last but not least, EE is an accomplice to computer abuse e-Crime resulting into distress, harassment and stalking (*paragraph (11)*) as stated under the Computer Misuse Act 1990 section 1 (1) (a) of Unauthorised access to computer material (Computer Misuse Act, 1990):

> *"A person is guilty of an offense if—(a) he causes a computer to perform any function with intent to secure access to any program or data held in any computer [F1, or to enable any such access to be secured]"*

## Decision

(18) As it is a recently reported e-Crime in the media (BBC, 2019), no court details and decisions were obtainable as of the month of March of the year 2019 because investigations are still ongoing. However, the Information Commissioner's Office was notified and it said it was illegal for individuals to access personal data without authorization as per the Data Protection Act and the General Data Protection Regulation (GDPR).

(19) Furthermore, since no legal decision has yet been made on this e-Crime, we can only refer to this case's relationship with other referenced case laws for possible decisions and judgments to this e-Crime. This will be covered in the *Opinions and Analysis* section below.

## Reasoning

(20) The General Data Protection Regulation (GDPR, 2018) ensures natural persons are protected with regards to the processing of their personal data. This means fundamental rights and freedoms of natural persons are protected as far as personal data processing is concerned. Violation of the regulations by any data controller can attract administrative fines in addition to corrective measures by the national authorities (in this case the Information Commissioner's Office for the UK). The fines or penalties must be effective, proportionate and dissuasive depending on the nature of individual cases. Depending on the outcome of the EE data breach investigations as well as the level of severity assessed, probable fines and penalties provisions are listed in Article 83(4) (5) of the General Data Protection Regulation (GDPR, 2018).

GDPR Article 83(4): "*administrative fines up to 10 000 000 EUR, or in the case of an undertaking, up to 2 % of the total worldwide annual turnover of the preceding financial year, whichever is higher*"

GDPR Article 83(5): "*administrative fines up to 20 000 000 EUR, or in the case of an undertaking, up to 4 % of the total worldwide annual turnover of the preceding financial year, whichever is higher*"

(21) In reaction to this e-Crime, the Information Commissioner's Office (ICO) said it is companies' obligations to ensure data was managed securely, and protected "against unauthorized or unlawful processing and against accidental loss, destruction or damage". This means in an event EE is proven to have failed to meet these obligations, it will be held accountable as per the Information Commissioner's Office guiding regulations and associated penalties as already hinted in *paragraph (20)*.

(22) Just like the General Data Protection Regulation (GDPR), the DPA (Data Protection Act, 2018) also makes provision about the processing of personal data for individuals. The Data Protection Act provisions for Information Commissioner's functions to market code of practice in relations to information handling. Incidences involving Data Protection Act violations such as those outlined in *paragraphs (14) (15)* immediately calls for Information Commissioner's Office interventions and investigations as a national authority. If investigation results show compliance failures, appropriate corrective measures and penalties are inevitable by the Information Commissioner's Office (ICO).

## Opinions

(23) Because of failure to follow internal policies as in *paragraph (12)*, Ms Bonafede's personal information was breached and she was exposed to stalking and damages such as those intended to cause her personal loss (financially); harassment; distress and privacy infringement. Therefore, Ms Bonafede is within her legal rights based on the Legal *Issues* presented to seek compensations in line with the damages faced and put EE accountable to ensure privacy and protection of all its customers' personal data.

(24) Following up from *paragraphs (20) (21) (22)*, it is evident EE should be subjected to penalties from the Information Commissioner's Office for negligence to protect against unauthorized access to personal data. This will ensure personal data for EE's over 35 million customers are well protected from all forms of unauthorized access due to privacy negligence in the future. It will also show how serious the issues relating to personal data breaches are, and positively encourage other organizations to enforce proper personal data processing.

## Analysis

*The significance of the case*

(25) This breach is a classic example demonstrating the requirement of General Data Protection Regulation (GDPR) Article 32 Security of processing data in all organizations business operations. This article of the GDPR highlights the need for the implementation of appropriate technical and organizational measures to ensure levels of security appropriate to the risks such as pseudonymization, encryption of personal data among others. It also directs the data controllers and processors how to ensure any natural person acting under the authority to follow appropriate guidelines when accessing subjects data while adhering to an approved code of conduct (GDPR, 2018).

(26) From this EE personal data breach case, the consequences of failing to follow set companies' internal policies and guidelines can be openly observed (*paragraph (12)*). Above all, the needs to set up controls to ensure personal data are well protected against unauthorized access by unauthorized personnel. Unauthorized personal data access could have all sorts of negative impacts with all levels of severity on the victims' lives. In this EE breach case, it could have led to other grave or violent crimes like impersonation due to identity selling, financial fraud, murder or even sexual harassments. Closely adhering to personal data protection guidelines is the only way of ensuring such scenarios never come to pass.

(27) This breach case further highlights the rampant negligence and increasing unprofessionalism in service businesses where it is possible staff could have been inadequately trained to professionally handle consumers' personal data. It therefore, raises the need by organizations to properly train and evaluate professional competencies of service support staff before allowing them to handle subjects' personal data and to ensure they comply with set rules, guidelines and regulations. Consulting staff training records would be the best way to determine the professional competence of the staff in executing their roles right. Regular refresher training for service support might be required as well to ensure professionalism and adherence to operational guidelines.

(28) Another notable significance of the EE breach case highlights' the values of customers' centricity and knowledge of customers' rights. As well, the responsibilities of companies in ensuring privacy are paramount so that similar incidences like Ms. Bonafede's which involves feedback like "*They didn't seem concerned*" are never heard of.

(29) As is recently common across many organizations, when it comes to information and cyber security they seem more prepared to defend against external intrusions and unauthorized access than are prepared against the risk of unauthorized data access by insiders such as staff in the organization. Therefore, this case raises such additional needs to equally invest in internal security and protection against unauthorized data access by unauthorized personnel within the organizations as well.

*Relationship to other referenced cases*

(30) A similar data breach case that has gone through court in the UK is that of (CR19 v The Chief Constable of the Police Service of Northern Ireland, 2014) in which the police force admitted liability in respect of a police officer's claim for negligence and breach of the Data Protection Act 1998 section 4 after his personal data and records had fallen into terrorists' hands.

(31) According to the *CR19 v Chief Constable of Northern Ireland* case proceeding, the police officer ("X") who represented himself at that hearing "failed to raise either the breach of Data Protection Act 1998 section 4 or the amount of any consequent compensation. The judge concluded that the damage established by X fell into the range of moderate psychiatric damage and awarded £20,000. The judge's compensatory award of £20,000 had taken account of the distress engendered by the breach of data protection and so was adequate. However, the officer was entitled to additional nominal damages of £1.00 to reflect the fact that there had been an admitted breach of section 4".

(32) In Ms. Bonafede's case after her personal data was breached, she was subjected to harassment, stalking, and distress. Meanwhile, in X's case, he developed post-traumatic stress disorder and a habit of excessive alcohol consumption after his personal data was breached. This makes both cases have similar effects that had a grave impact and affected the health of their victims.

(33) Both the EE breach and X's case involved negligence and breach of statutory duty by the respective data controllers and processors. Since the negligence and breach of statutory duty had similar effects on their victims, compensating Ms. Bonafede would only be the most favorably fair option considering the (CR19 v The Chief Constable of the Police Service of Northern Ireland, 2014) case law. However, the GDPR of May 2018 will now have to be considered in making the judgment as well.

(34) Another reference case involving privacy breach similar to Ms. Bonafede's is *Gulati v MGN Ltd case* (Shobna Gulati & ors v MGN Limited, 2015). This is a privacy breach by phone hacking of phones belonging to public figures by a newspaper company MGN Ltd to gain access to confidential and private information. It is a claim based fundamentally on infringements of privacy rights. The difference with this case is that the defendant deliberately participated in the privacy breach and so all elements of the negligence of duty are dismissible. However, just as highlighted in *paragraph (32)* with X and Ms. Bonafede's circumstances, this case equally had some negative impacts on the victims' lives that resulted in compensation demands by the claimants. Similar guidance on damages payable in this and other phone hacking cases can be applied in Ms. Bonafede's case law.

(35) In this *Gulati v MGN Ltd* case, the privacy right to which the UK was obliged to give effect under the Human Rights Act 1998 was reflected in, though not itself created by, *European Convention on Human Rights (ECHR) Article 8*. The decisions held in this case can be used to substantiate on *paragraph (16)* when making a judgment.

(36) Considering the TalkTalk Company hacking case (Regina v Connor Douglas Allsopp, 2019) which was presented before Lord Justice Flaux, Mr Justice Sweeney and Mr Justice Soole on Wednesday 30th January 2019; a 22-year-old appellant and his co-accused, Matthew Hanley, pleaded guilty on re-arraignment to count 3 (causing a computer to perform a function to secure or to enable unauthorized access to a program or data, contrary to section 1(1) of the Computer Misuse Act 1990). This is the same law that was violated in the EE breach when EE gained unauthorized access to Ms. Bonafede's phone as is presented in *paragraph (17)*. It, therefore, makes EE susceptible to the same sentence as of the TalkTalk hacking case law.

*EE data breach e-Crime's place in history*

(37) According to the Office of National Statistics Crime in England and Wales: year ending September 2018, there was a 13% decrease in unauthorized access to personal information (including hacking). There were 470,000 recorded incidences between October 2017 to September 2018 compared to the figure of 542,000 between October 2016 and September 2017 (Office of National Statistics, 2019). However, there have still been many high profile data breaches that have affected people in the UK such as the breach cases from the British Airways, NHS, Marriott, Ticketmaster, Butlin's, Dixons Carphone among others (British Telecommunications, 2018). Almost all these reported cases at least have one thing in common, that is involving an external cyber hack or breach from outside the organizations.

(38) The mobile and telecommunications industry has also had numerous reported notable data breaches involving unauthorized access or hacks. A few common examples of such hacks or breaches include TalkTalk (BBC, 2018), Vodafone (The Guardian, 2015) and now this EE data breach. Much as there has been numerous undisclosed unauthorized data access to personal data by internal mobile and telecommunications staff, publicly exposed Ms Bonafede's personal data breach by EE is the first to have caused such a public stir since the official enactment of the General Data Protection Regulation (GDPR) of May 2018 in the UK. It was a social media engagement that forced EE to take action and start handling the breach more seriously. It can be presumed that had it not been for the social media engagements, Ms. Bonafede's personal data breach investigation by EE would still be dragging.

*The value of the current laws in place*

(39) Data Protection Directives and related rules 'could best be regarded as principles of good governance', as these rules were not framed as relating to the human rights of the data subject, but rather focussed on the procedural obligations of controllers (Sloot, 2014) (McDermott, 2017).

(40) The current laws in place, therefore, protect fundamental rights and freedoms of natural persons and in particular their right to the protection of personal data. The laws ensure all citizens are protected from privacy and data breaches in today's data-driven world by reshaping how data is managed by sectors and businesses (GDPR, 2018).

## About the Author

*Samuel Cris Ayo* received a BEng degree in Electronic Information Engineering from China University of Geosciences, Wuhan in 2011 and also holds a Professional Marketing certification with specialty to Digital Marketing from the Chartered Institute of Marketing, United Kingdom. He is currently working towards a Masters degree in Information Security and Digital Forensics in the School of Architecture, Computing, and Engineering at University of East London, United Kingdom. For 5 years since 2012, he worked as Products Service Manager in the Consumer Business Group and 1 year as a Projects supply manager with Huawei Technologies. Current interests revolve around digital forensics, cybersecurity, security and risks management.